\documentclass[aps,prd,twocolumn,times,graphicx,amssymb,tighten,showpacs,groupedaddress,superscriptaddress,floatfix]{revtex4-1}

\usepackage{graphicx,amssymb,amsmath}
\usepackage[dvips]{xcolor}
\usepackage{hyperref}
\usepackage{amsmath}
\usepackage{amssymb}
\newcommand{\be}{\begin{equation}}
\newcommand{\ee}{\end{equation}}

\begin{document}
\title{Inclusive electron scattering within the SuSAv2-MEC approach}

\author{G.D.~Megias}
\affiliation{Departamento de F\'{i}sica At\'omica, Molecular y Nuclear, Universidad de Sevilla, 41080 Sevilla, Spain}

\author{J.E.~Amaro}
\affiliation{Departamento de F\'isica At\'omica,  Molecular y Nuclear and Instituto Carlos I de F\'isica Te\'orica y Computacional, Universidad  de  Granada, 
18071 Granada, Spain}

\author{M.B.~Barbaro}
\affiliation{Dipartimento di Fisica, Universit\`{a} di Torino and INFN, Sezione di Torino, Via P. Giuria 1, 10125 Torino, Italy}

\author{J.A.~Caballero}
\affiliation{Departamento de F\'{i}sica At\'omica, Molecular y Nuclear, Universidad de Sevilla, 41080 Sevilla, Spain}

\author{T.W.~Donnelly}
\affiliation{Center for Theoretical Physics, Laboratory for Nuclear Science and Department of Physics, Massachusetts Institute of Technology, Cambridge, 
Massachusetts 02139, USA}

\date{\today}
\begin{abstract}
We present our recent progress on the relativistic modeling of electron-nucleus reactions
and compare our predictions with inclusive $^{12}$C ($e,e'$) experimental data in a wide kinematical region. 

The model, originally based on the superscaling phenomenon shown by electron-nucleus scattering data, has recently been improved through the inclusion of
Relativistic Mean Field theory effects that take into account the enhancement of the quasielastic transverse scaling function compared with its longitudinal counterpart. 
In this work we extend the model to include the complete 
inelastic spectrum -- resonant, non-resonant and deep inelastic scattering (DIS). 
We also discuss the impact of meson-exchange currents (MEC)
through the analysis of two-particle two-hole 
contributions to electromagnetic response functions 
evaluated within the framework of the relativistic Fermi gas, considering for the first time not only the transverse but also the longitudinal channel.
The results show quite good agreement 
with data over the whole range of energy transfer, including the dip region between the quasielastic peak and the $\Delta$ resonance. 
\end{abstract}


\maketitle

\section{Introduction}
\label{intro}

One of the challenging goals of current neutrino oscillation experiments is a proper and precise description of neutrino-nucleus scattering 
at intermediate energies (from few hundred MeV to few GeV). Particular emphasis is placed on the evaluation of effects linked to the nuclear
structure involved in the analysis of experiments. In recent years, several models, originally developed to study electron-nucleus scattering,
have been further extended to the description of neutrino-nucleus cross 
sections~\cite{Amaro:2004bs,Amaro:2006if,Martini:2009aa,Meucci:2003,Meucci:2004,Nieves:2004,Benhar2010:aa,Lalakulich2012:aa}.
These models are required to provide a precise enough description of electron scattering data before they can be applied to neutrino reactions.
In some cases, such as the simple and commonly-used relativistic Fermi gas model (RFG), they fail to reproduce both inclusive electron scattering in 
the quasielastic (QE) regime as well as recent measurements of QE neutrino and antineutrino scattering cross sections. This is connected to
the approaches assumed by the specific nuclear models and, more importantly, with the simplified description of the reaction mechanism that in
most of the cases is based on the impulse approximation (IA) with additional non-relativistic reductions. Hence a proper evaluation of the effects
introduced by final-state interactions (FSI) and mechanisms beyond the IA, such as
nuclear correlations and two-particle two-hole excitations, are needed.
In this context, a consistent and complete description of the electron scattering cross section that includes not only the QE regime, but also
regions at higher energy transfer (nucleon resonances, inelastic spectrum), is essential
for the analysis of current neutrino oscillation experiments. This 
provides a critical baseline for the validation of theoretical neutrino-nucleus interaction models.

In recent years, the scaling~\cite{Day:1990} and superscaling properties~\cite{super,super1} of electron-nucleus interactions have been 
analyzed in detail and used to construct a semi-phenomenological model for lepton-nucleus scattering~\cite{Amaro:2004bs}. 
This model, denoted as SuperScaling Approach (SuSA)~\cite{super,super1,Chiara1}, 
assumes the existence of universal scaling functions for electromagnetic and weak interactions. 
The general procedure adopted in
this analysis consists of dividing 
the ($e,e'$) experimental cross section by an appropriate single-nucleon one to obtain a reduced cross section. When this is plotted as a function of 
the ``scaling'' variable ($\psi$), itself a function of the energy ($\omega$) and momentum transfer ($q$), some particular properties emerge. 
Specifically, analyses of inclusive $(e,e')$ data have shown that at energy transfers below the QE peak, the reduced cross section is 
largely independent of the momentum transfer, which is called scaling of first kind, and of the nuclear target, which is defined as scaling 
of second kind. This simultaneous occurrence of scaling of both kinds is denoted as superscaling. At higher energies, above the QE peak, both 
kinds of scaling are shown to be violated as a consequence of the contributions introduced by effects beyond the impulse approximation (IA), 
such as 
meson-exchange currents (MEC) and inelastic scattering. 
An extension of the scaling formalism, originally introduced to describe the QE domain, to the region of the $\Delta$ resonance and the complete inelastic 
spectrum -- resonant, non-resonant and deep inelastic scattering (DIS) -- has also been proposed in \cite{Barbaro:2003ie,Maieron:2009an,Ivanov:2016Delta}.

Recently we have developed an improved version of the superscaling model, called SuSAv2~\cite{Gonzalez-Jimenez:2014eqa}, that incorporates 
relativistic mean field (RMF) effects \cite{Caballero:2005sj,Caballero:2006wi,Caballero:2007tz} in the longitudinal and transverse nuclear responses, 
as well as in the isovector and isoscalar channels independently. Note that the RMF model leads to a natural enhancement of the transverse response 
through RMF effects without resorting to inelastic processes or two-particle emission via MEC. The RMF works properly at low to intermediate values of the momentum transfer, $q$. However, because of the strong
energy-independent scalar and vector potentials involved, the RMF does less well at higher values of $q$, where the Relativistic Plane Wave Impulse
Approximation (RPWIA) gives better predictions. Hence both regimes are incorporated in SuSAv2 by making use of a reasonable "blending" function~\cite{Gonzalez-Jimenez:2014eqa}.

While the original SuSAv2 was based exclusively on the IA, and used to describe the QE domain, in this work the model is extended to the inelastic spectrum.
Following previous studies on the inelastic RFG modeling \cite{Barbaro:2003ie} we achieve this goal by employing 
phenomenological fits to the single-nucleon inelastic structure functions.

Ingredients beyond the IA, namely, 2p-2h MEC effects, have been shown to play an important role in the ``dip'' region between the QE and the $\Delta$ peaks.
In this work the SuSAv2 model also incorporates contributions in both longitudinal and transverse reaction channels arising from
2p-2h  states  excited  by  the  action  of  electromagnetic, purely  isovector  meson-exchange  currents within a fully relativistic framework 
(see \cite{DePace:2003xu,Simo:2014wka,Simo:2014esa,Megias:2014qva} for details). Therefore, the new ``SuSAv2-MEC'' predictions can be compared with data for
very different kinematical situations, covering the entire energy spectrum. The accordance between theory and data gives us
confidence in the
extension of the model and its validity when applied to recent neutrino oscillation experiments where all the different kinematical regions may contribute and, in particular,
effects linked to 2p-2h MEC have been claimed to be essential in order to reproduce the neutrino-nucleus scattering cross sections \cite{Martini:2009aa,Nieves:2011pp,Megias:2014qva}.

This paper is organized as follows. In  Sect.~II we briefly introduce the formalism for QE and inelastic lepton-nucleus reactions and describe how the MEC 
have been computed.
In Sect.~III we compare our predictions with inclusive ($e,e'$) experimental data in a wide kinematical region. The analysis is presented for the cross 
sections 
paying a special attention to the relevance of the RMF and RPWIA effects at different kinematics. 
Finally, in Sect. IV we show the conclusions of our analysis, including some remarks related to studies of neutrino reactions with nuclei.

\section{General formalism: the model}

\subsection{SuSAv2 in the QE region}

Following the Rosenbluth prescription~\cite{Rosenbluth:1950}, the double differential ($e,e'$) inclusive cross section (differential with respect to the electron scattering angle $\Omega_e$ and the transferred energy $\omega$) is given as the sum of two response functions corresponding to the longitudinal, $R_L$, and transverse, 
$R_T$, channels ($L$ and $T$ refer to the direction of the transferred momentum, $q$),
\begin{equation}\label{ecsec}
 \displaystyle \frac{d^2\sigma}{d\Omega_e d\omega}=\sigma_{Mott}(v_LR_L+v_TR_T) ,
\end{equation}
where $\sigma_{Mott}$ is the Mott cross section and the $v$s
are kinematical factors that involve leptonic variables (see \cite{Day:1990} for explicit expressions). Assuming
charge symmetry, these two channels can be decomposed as a sum of the isoscalar ($T=0$) and
isovector ($T=1$) contributions. In terms of the scaling functions the nuclear responses
are
\begin{eqnarray}\label{eresp}
 \displaystyle R_{L,T} (q,\omega)&=&\frac{1}{k_F}\Bigl[f_{L,T}^{T=1}(\psi')G_{L,T}^{T=1}(q,\omega)\nonumber\\
\displaystyle &+&f_{L,T}^{T=0}(\psi')G_{L,T}^{T=0}(q,\omega)\Bigr] ,
\end{eqnarray}
where  $k_F$ is the Fermi momentum and the $f$s are the scaling functions, that only depend on the scaling variable $\psi'$. This scaling variable depends on $q$, $\omega$ and on the energy shift, 
$E_{shift}$, needed in order to have the corresponding scaling function peak located at $\Psi'=0$, as described in \cite{Gonzalez-Jimenez:2014eqa}.

The functions $G_{L,T}^{T=0,1}$ are defined as the isoscalar and isovector 
responses of a moving nucleon and include relativistic corrections arising from the presence of the medium. Their explicit expressions, not reported here for the sake of brevity, can be found in \cite{Gonzalez-Jimenez:2014eqa,Day:1990}.

\begin{figure}[H]
\begin{center}
\vspace{-1.8cm}\includegraphics[scale=0.425, angle=270]{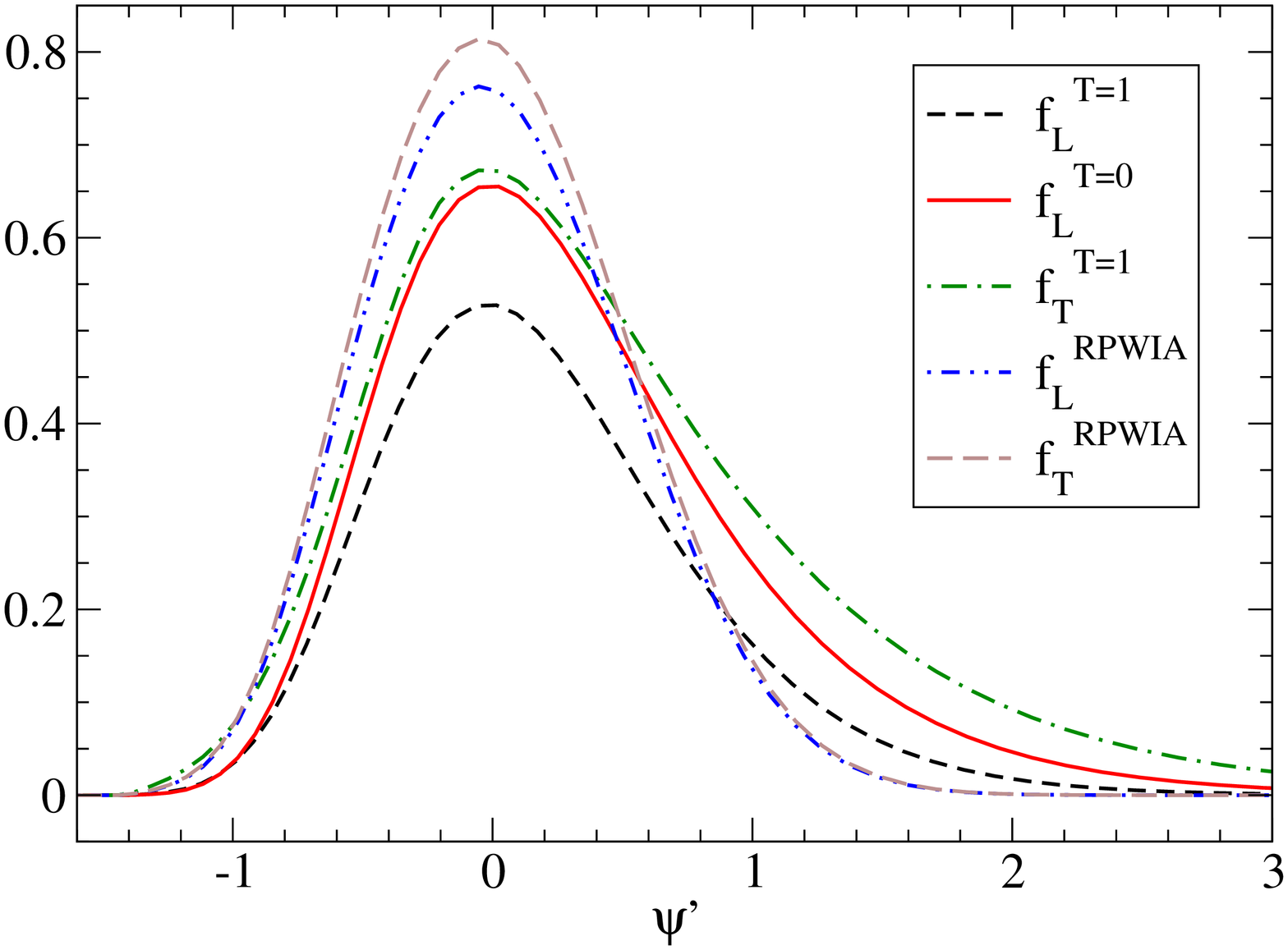}
\end{center}\vspace{-1.0cm}
\caption{(Color online) Reference scaling functions in the SuSAv2 model.}\label{susav2ref}
\end{figure}

In Fig.~\ref{susav2ref} we present the scaling functions of relevance for electron-nucleus reactions, based on results from~\cite{Gonzalez-Jimenez:2014eqa}. 
Some basic conclusions emerge from the analysis of the scaling functions in the RMF and RPWIA models. First, the
two models differ in the treatment of the final state. Whereas the RPWIA describes the outgoing nucleon as a relativistic plane wave, the RMF takes into account 
FSI between the outgoing nucleon and the residual nucleus using the same mean field as considered for the bound nucleon. This leads to a violation of the
so-called zeroth-kind scaling, that is, the RMF transverse and longitudinal scaling functions differ from each other, the former being larger  by an amount of the order of $20\%$. This is directly linked to the distortion introduced by FSI in the lower components of the 
outgoing nucleon Dirac wave functions. Secondly, it is also noteworthy that the tail exhibited by the scaling function at large values of 
$\omega$ is significantly higher and more extended in the transverse channel. On the contrary, the results obtained within the RPWIA show that the two types of 
scaling functions are roughly the same, having a shape that is much more symmetric, {\it i.e.,} lacking the long tail extending to large values of $\omega$. 

In spite of the merits of the RMF description,
a particular drawback of the RMF concerns its dependence upon the momentum transfer $q$: indeed the RMF peak position keeps growing with $q$, thus making questionable the validity of the model at very high $q$. 
In fact,
the large kinetic energy of the outgoing nucleon at very high $q$
should make the FSI effects negligible. Thus, it would be desirable 
that the RMF scaling functions approach the RPWIA ones for increasing momentum transfer~\cite{Gonzalez-Jimenez:2014eqa}. This was a basic motivation in the
development of a new SuperScaling Approach as a combination of RMF and RPWIA scaling functions where the first dominates at low to intermediate $q$ 
and the latter at high $q$. This implies that the scaling functions in Eq.~(\ref{eresp}) should be replaced by linear combinations of 
RMF-based ($\tilde{f}_{L,T}$) and RPWIA ($\tilde{f}^{RPWIA}_{L,T}$) scaling functions: 

\begin{eqnarray}\label{scf}
 \mathcal{F}_L^{T=0,1}&\equiv& \cos^2\chi(q)\tilde{f}_L^{T=0,1}+\sin^2\chi(q)\tilde{f}_L^{RPWIA} \nonumber\\
\mathcal{F}_T&\equiv& \cos^2\chi(q)\tilde{f}_T+\sin^2\chi(q)\tilde{f}_T^{RPWIA} \, , \nonumber\\
\end{eqnarray}
where $\chi(q)$ is a $q$-dependent angle given by 
\begin{equation}\label{xi2}
\displaystyle\chi(q)\equiv\frac{\pi}{2}\left(1-\left[1+e^{\left(\frac{\left(q-q_0\right)}{\omega_0}\right)}\right]^{-1}\right)
\end{equation}
and the transition between RMF and RPWIA behaviors occurs at intermediate $q$-values ($q_0$) in a region of width $\omega_0$, which is fixed at 200 MeV. Notice that the separation 
into isoscalar ($T=0$) and isovector ($T=1$) contributions is only taken into account for the RMF longitudinal function as in the transverse component the isoscalar contribution is negligible. In contrast, for the RPWIA longitudinal and transverse scaling functions, the isovector and isoscalar contributions collapse into a single curve.

The electromagnetic response functions are now defined as:
\begin{eqnarray}\label{eresp1}
 \displaystyle R_{L} (q,\omega)&=&\frac{1}{k_F}\Bigl[\mathcal{F}_{L}^{T=1}(\psi')G_{L}^{T=1}(q,\omega)\nonumber\\
\displaystyle &+&\mathcal{F}_{L}^{T=0}(\psi')G_{L}^{T=0}(q,\omega)\Bigr] \\
 \displaystyle R_{T} (q,\omega)&=&\frac{1}{k_F}\mathcal{F}_{T}^{T=1}(\psi')\Bigl[G_{T}^{T=1}(q,\omega)\nonumber\\
\displaystyle &+& G_{T}^{T=0}(q,\omega)\Bigr] \, .
\end{eqnarray}

Thus, the transition between the two models depends on the particular kinematics involved, namely, on the momentum transfer $q$. Accordingly,
the transition parameter, $q_0$, is expected to increase with $q$ in such a way that the RMF contribution will be dominant at low kinematics 
whereas the RPWIA one starts to be relevant at higher energies. 
 Therefore we introduce a dependence of the parameter $q_0$ on the momentum transfer $q$
that determines the relative RMF and RPWIA contributions at different kinematics.

The particular procedure to determine the $q_0$-behavior with $q$ is in accordance to the best fit to a large amount of $(e,e')$ experimental data, 
covering from low to high $q$-values ($q$:239-3432 MeV/c). The method applied is based on a reduced-$\chi^2$ analysis of the data sets. This analysis is performed 
in conjunction with the inelastic one, taking into account also the 2p-2h MEC contributions, and will be detailed in Sect.~\ref{2d}.

\subsection{Inelastic electron-nucleus scattering in the SuperScaling Approach}

The general formalism describing inclusive inelastic electron-nucleus scattering in the SuperScaling Approach has been presented in previous 
work~\cite{Barbaro:2003ie}. Here we consider a more sophisticated description of the lepton-nucleus reactions 
via RMF and RPWIA ingredients (SuSAv2 model). 
The hadronic tensor for inelastic processes can be written in the form~\cite{Barbaro:2003ie}: 
\begin{eqnarray}
\displaystyle W^{\mu\nu}_{inel}(\mbox{\bf q},\omega)=\frac{3\mathcal{N}}{4\pi k_F^3}\int_F d\mbox{\bf h}\frac{m_N}{\bar{E}_h}w^{\mu\nu}_{inel}(H,Q,\omega+\bar{E}_h)
\, \nonumber \\
\end{eqnarray}
with $k_F$ the Fermi momentum and $H$ and $\overline{E}_h=\sqrt{h^2+m_N^2}$ the 4-momentum and energy of the on-shell nucleon in the nucleus attached to the virtual photon.
The inelastic longitudinal and transverse responses functions, given by specific components of the hadronic tensor: 
$R^L_{inel}=W^{00}_{inel}$ and $R^T_{inel}=W^{11}_{inel}+W^{22}_{inel}$, can be expressed as
\begin{eqnarray}
\displaystyle R ^{L,T}_{inel}(\mbox{\bf q},\omega)=\frac{\mathcal{N}}
{\eta_F^3\kappa}\xi_F\int^{1+2\lambda-\varepsilon_S}_{\mu_{thresh}}d\mu_X\mu_X\mathcal{F}_{L,T}(\psi'_X)U_{L,T} \, , \nonumber \\
\end{eqnarray}
where we have introduced the dimensionless variables $\kappa = q/2m_N$, $\xi_F=\sqrt{1+(k_F/m_N)^2}-1$ and $\varepsilon_S=E_S/m_N$ with 
$m_N$ the nucleon mass and $E_S$ the separation energy (see~\cite{Barbaro:2003ie} for details). 
The parameter $\mu_X$ is the dimensionless invariant mass and $\mu_{thresh}$ refers to the pion-production threshold. 
The terms $\mathcal{F}_{L,T}$ are the inelastic scaling functions which exhibit the same structure as in Eq.~(\ref{scf}), but using the inelastic scaling 
variable $\psi'_X$. Finally, the functions $U_{L,T}$, firstly introduced in~\cite{Barbaro:2003ie}, depend on the single-nucleon inelastic structure functions 
$w_{1,2}$ which are described in our case by using empirical fits of the inelastic electron-proton and electron-deuteron cross sections~\cite{Bosted1,Bosted2}.

As already commented on for the QE case, the determination of the RMF/RPWIA transition parameter ($q_0$) in the inelastic regime also depends on the 
particular kinematics involved and it will be discussed in detail in Sect.~\ref{2d}.
\subsection{Electromagnetic 2p-2h MEC contributions}

\begin{figure}[H]
\begin{center}
\vspace{-1.8cm}\includegraphics[scale=0.425, angle=270]{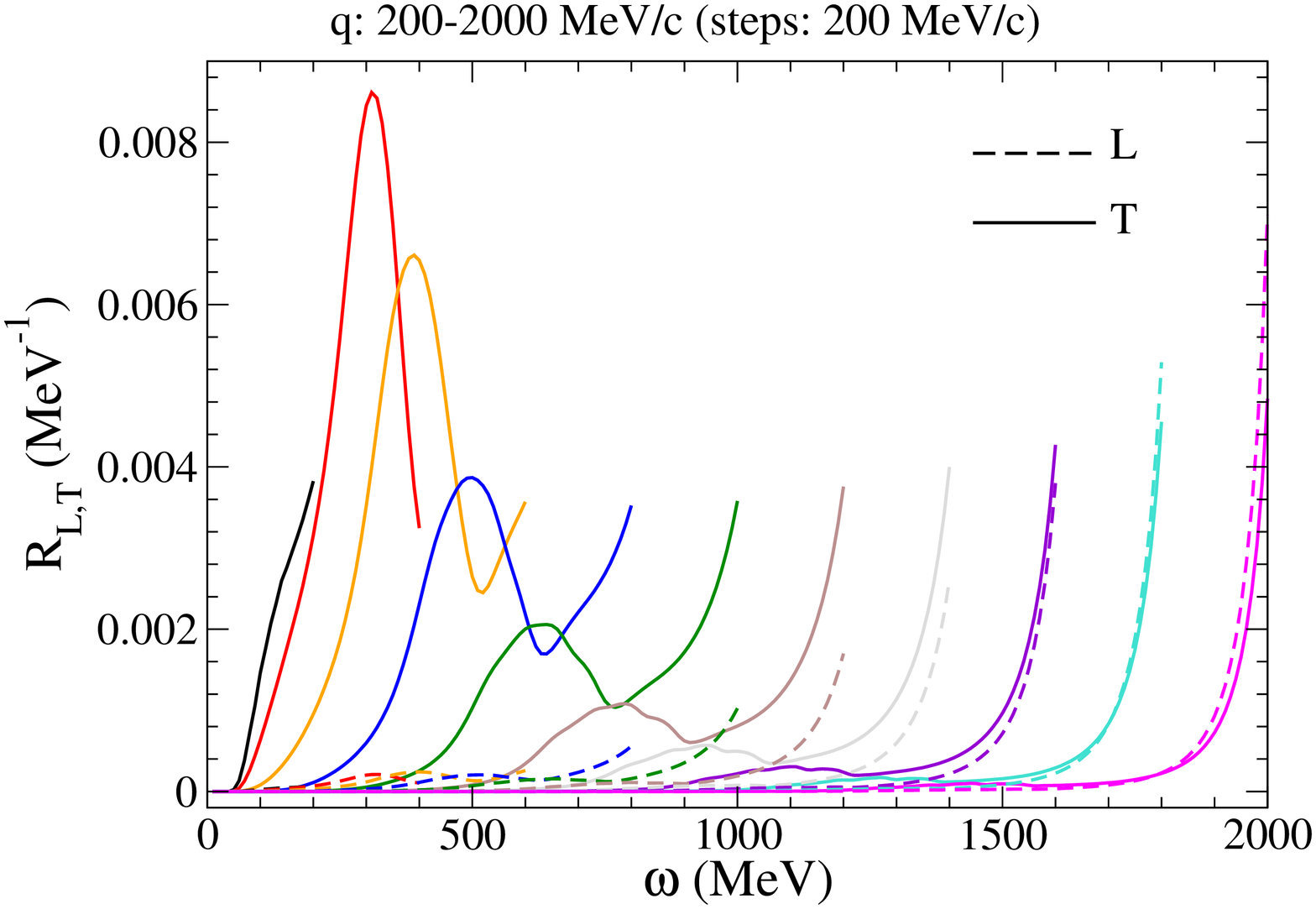}
\end{center}\vspace{-1.0cm}
\caption{(Color online) Comparison between 2p-2h MEC $R_L$ and $R_T$ response functions versus $\omega$. The  curves  are  displayed  from  left  to  right  in  steps  of
$q=$200 MeV/c.}\label{mec1}
\end{figure}

The evaluation of the 2p-2h pionic MEC contributions is
performed within the RFG model in which a fully Lorentz covariant calculation of the MEC can be performed (see \cite{Megias:2014qva,DePace:2003xu,Simo:2014wka}). Although MEC
clearly dominate in the transverse channel, our present study includes 
also for the first time MEC contributions in the longitudinal sector. In Fig.~\ref{mec1}
we present the separate 2p-2h MEC responses in the two channels. As
shown, the transverse sector clearly dominates up to $q\sim1800$
MeV/c, while the $L$ and $T$ contributions are of the same order for
larger values of the momentum transfer.  However, note that 
the kinematics where the MEC give the largest contribution to the cross section corresponds to $q\lesssim1000-1500$ MeV/c, as stated in~\cite{Megias:2014qva}.

\begin{figure}[H]
\begin{center}
\vspace{-1.8cm}\includegraphics[scale=0.425, angle=270]{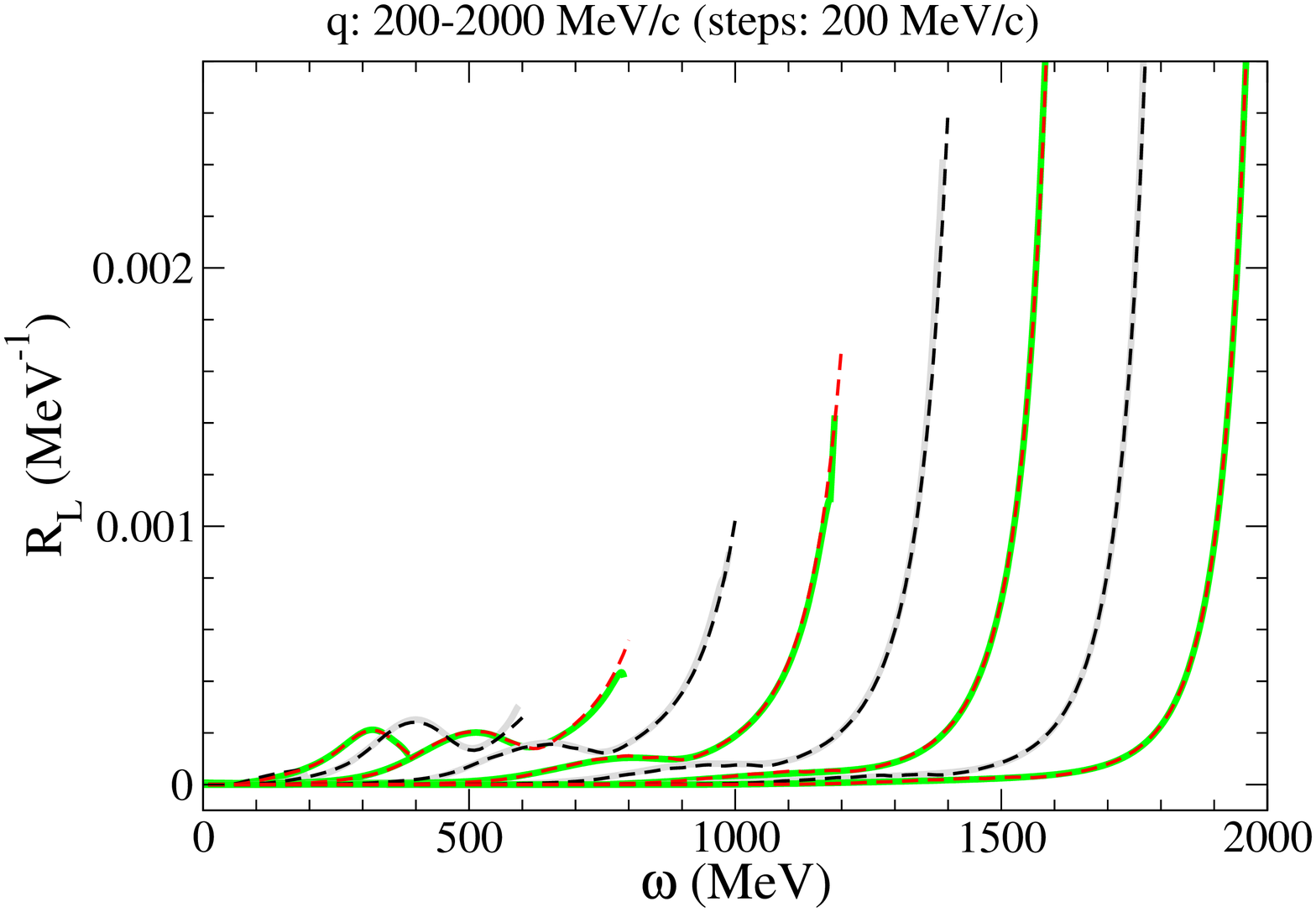}
\end{center}\vspace{-1.0cm}
\caption{(Color online) Comparison between the longitudinal 2p-2h MEC response functions (dashed lines) and the parameterized ones (thick solid lines) versus $\omega$. 
The curves are displayed from left to right in step of $q=$200 MeV/c.} \label{mec2}
\end{figure}

As discussed in previous work  
\cite{DePace:2003xu,DePace:2004xu,Amaro:2010iu,Simo:2014esa,Simo:2014wka}, 
relativity is an essential ingredient in the analysis 
of 2p-2h processes at momentum transfers above 500 MeV/c. 
At these $q$-values, the static approximation used for the $\Delta$ propagator
in the non-relativistic calculations of $2p-2h$ transverse response function \cite{Amaro:1994fx} fails to explain the ``dip'' region.
A fully relativistic calculation of the 2p-2h MEC response functions 
in the RFG model requires one to compute the spin-isospin traces of all the many-body 
MEC diagrams.
This involves the analytical calculation of more than 100,000 terms some of which involve subsequent numerical seven-dimensional integrations.
This makes the computation highly non-trivial.
In order to reduce the computational time as well as to ease the implementation of the results in Monte Carlo generators used in the analysis 
of neutrino experiments, where a wide range of kinematic conditions --- momentum and energy transfers --- are involved, we make use of a parametrization of the MEC responses.
The parametrization form employed for the transverse electromagnetic response was analyzed in~\cite{Megias:2014qva}. In the present work, we follow a similar
procedure to get a description for the longitudinal one. As shown in Fig.~\ref{mec2} the 2p-2h MEC longitudinal response function is reproduced with a high 
accuracy, a result very similar to the situation already presented in the transverse channel (see~\cite{Megias:2014qva}).
Notice that no approximations are involved in the present calculation. The MEC parametrization considered here takes care of the complete relativistic
calculation, making it suitable to be applied at very high values of the momentum and energy transfers.
\subsection{Determination of $q_0$ parameters}\label{2d}

\begin{figure}[H]
\begin{center}
\vspace{-1.8cm}\includegraphics[scale=0.425, angle=270]{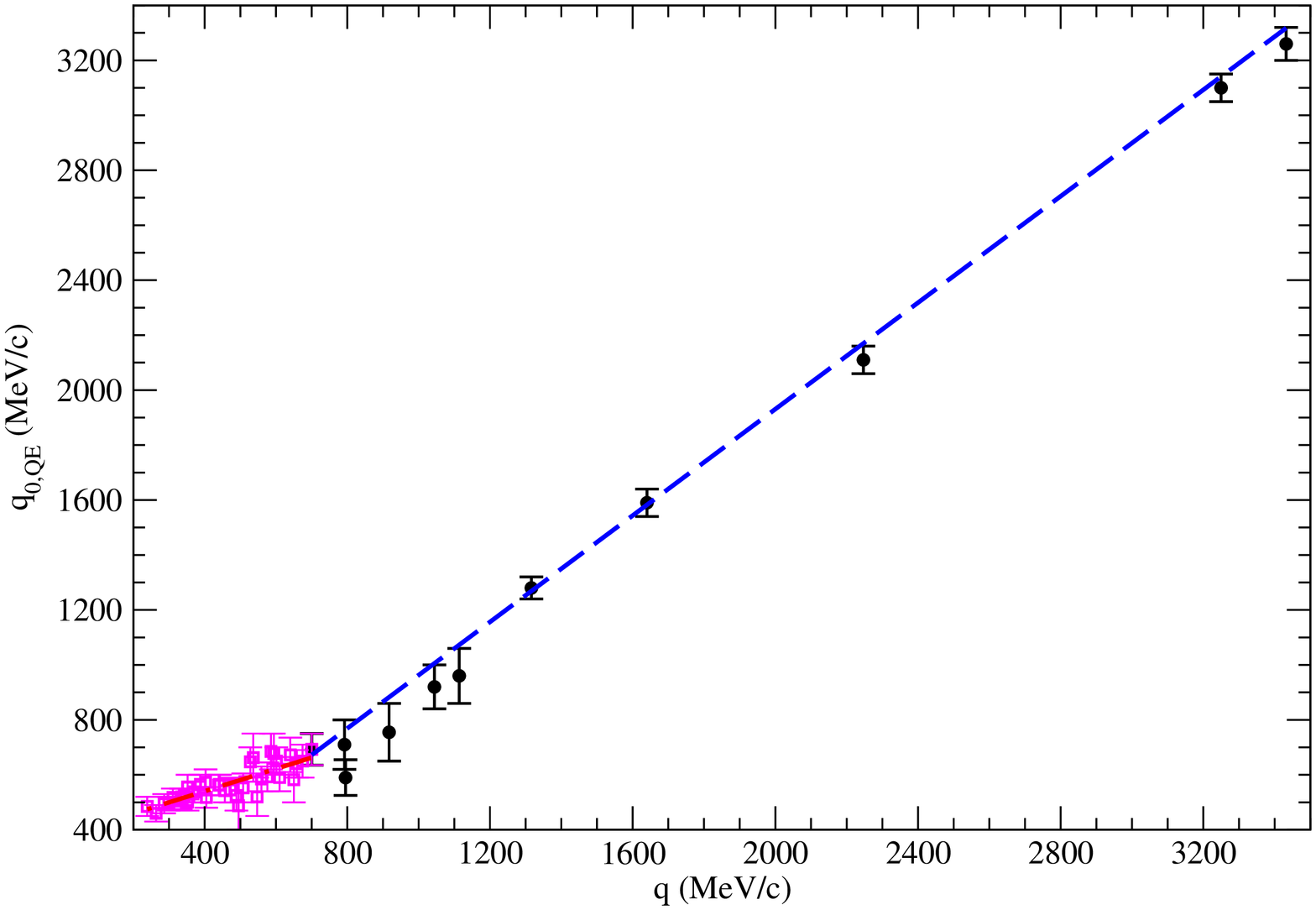}\\
\vspace{-0.8cm}
\includegraphics[scale=0.425, angle=270]{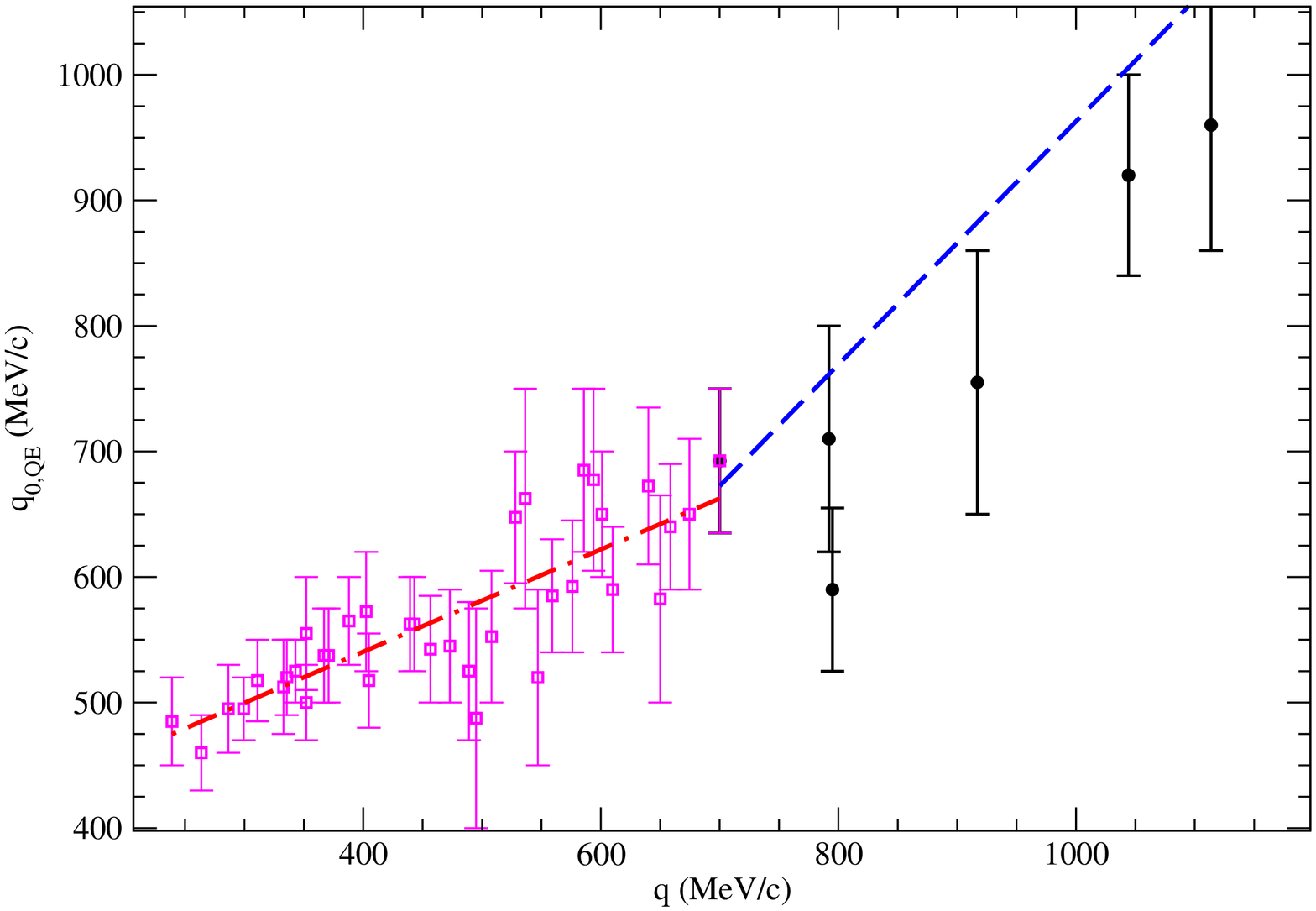}\\
\vspace{-0.8cm}
\includegraphics[scale=0.425, angle=270]{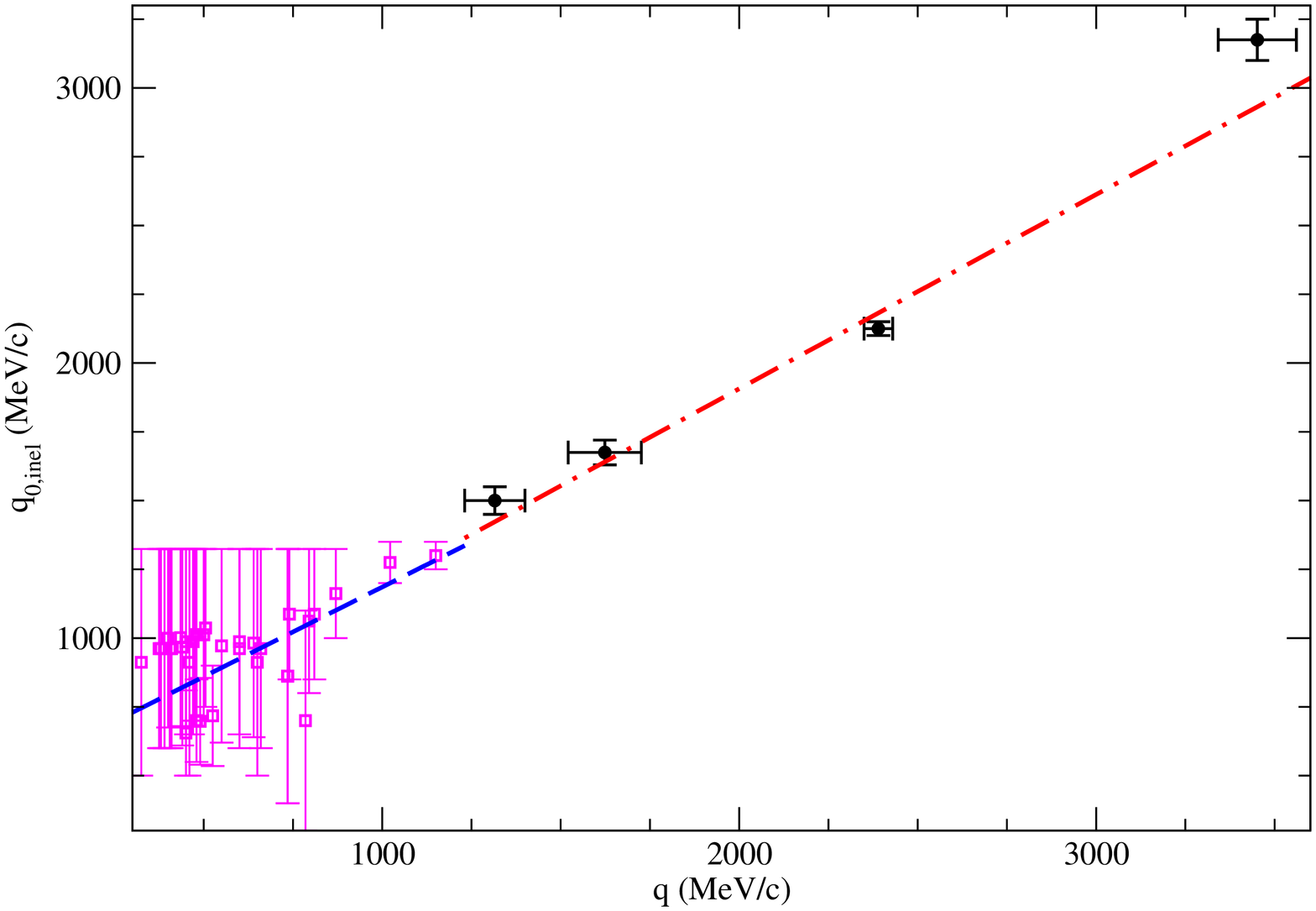}\\
\end{center}\vspace{-1.0cm}
\caption{(Color online) Parametrization of $q_{0,QE}$ in terms of $q$ (top and middle panels). Data points represent the $q_0$ value that best fits each case. 
  Parametrization of $q_{0,inel}$ in terms of $q$ (bottom panel).
}\label{q0param}
\end{figure}

The procedure to determine the RMF/RPWIA transition in the SuSAv2
model in both QE and inelastic regimes is based on the analysis of the
($e,e'$) data in a wide kinematical region. The transition parameter,
$q_0$ (see Eq.~(\ref{xi2})), must exhibit a dependence on the
particular kinematics involved in such a way that at higher energies,
which imply higher momentum transfers, the RPWIA contribution is more
relevant than the RMF one, whereas the opposite occurs at lower
energies. With these assumptions, we perform a $\chi^2$ analysis of
the electron-nucleus experimental data which is first focused on the
QE region ($q_0^{QE}$) and after that extended to the inelastic domain
($q_0^{inel}$). In the whole analysis we take into account the SuSAv2
model for both QE and inelastic regimes as well as the 2p-2h MEC
calculations. After analyzing the experimental data set, we get the
$q_0^{QE}$ and $q_0^{inel}$ parameters as functions of
$q$. Figure~\ref{q0param} illustrates the behavior of both parameters,
$q_0^{QE}$ (top and middle panels) and $q_0^{inel}$ (bottom).  
The data points and their error bands represent the values of the parameters that best fit the data
at different kinematics (within a $\sim10\%$ in the $\chi^2$ minimum).
As shown, $q_0^{QE}$ increases moderately with $q$ at low to intermediate
values whereas the slope goes up significantly at higher kinematics
($q\gtrsim700$ MeV/c).

This suggests the following parametrization:
\begin{equation}\label{q0qe}
q_0^{QE}(q)= \left\{ \begin{array}{cc} A+B q, & q < q_1 \\
                                      C+D q, & q> q_1
                     \end{array}
            \right. 
\end{equation}
with $q_1=700$MeV/c, $A=377.629$ MeV/c, $B=0.407$, $C=-5.322$ MeV/c and $D=0.968$. Imposing continuity of the above function we are
left with three free parameters, $A, B, C$, in the fit.  

A similar parametrization  is found for  
$q_0^{inel}(q)$, but in this case  only one linear function is used for the whole region of $q$ explored.
\begin{equation}\label{q0inel}
q_0^{inel}(q)= A'+B' q
\end{equation}
with $A'=494.439$ MeV/c and $B'=0.706$.

Finally, it is also worth
mentioning that an even better agreement with the ($e,e'$) data could be
achieved by employing a non-linear fit of the $q_0$ parameters as well

as including a dependence on the incident energy ($E_i$) or the
scattering angle ($\theta_e$) in the transition parameters
($q_0,\omega_0$); however, the simpler assumptions made in this work are felt to be adequate for our purposes. 
%

\section{Results}

In this section we present our results for $^{12}$C($e,e'$) cross sections.
In the following we adopt the 
Bosted and Christy parametrization for the single-nucleon inelastic structure functions \cite{Bosted1,Bosted2} which describes DIS, resonant and 
non-resonant regions. For the QE regime, we employ the electromagnetic form factors of the extended Gari-Krumpelmann (GKex) 
model~\cite{Lomon1,Lomon2,Crawford1}. The sensitivity of the QE results to the different parametrizations has been discussed 
in \cite{Megias:2013aa}. Additionally, for the Fermi momentum we employ the values obtained in~\cite{Chiara1}, namely $k_F=228$ MeV/c for $^{12}$C.

\subsection{Differential cross sections}

\begin{figure}[H]
\begin{center}\vspace{-1.9cm}
\includegraphics[scale=0.99, bb=50 109 594 800, clip]{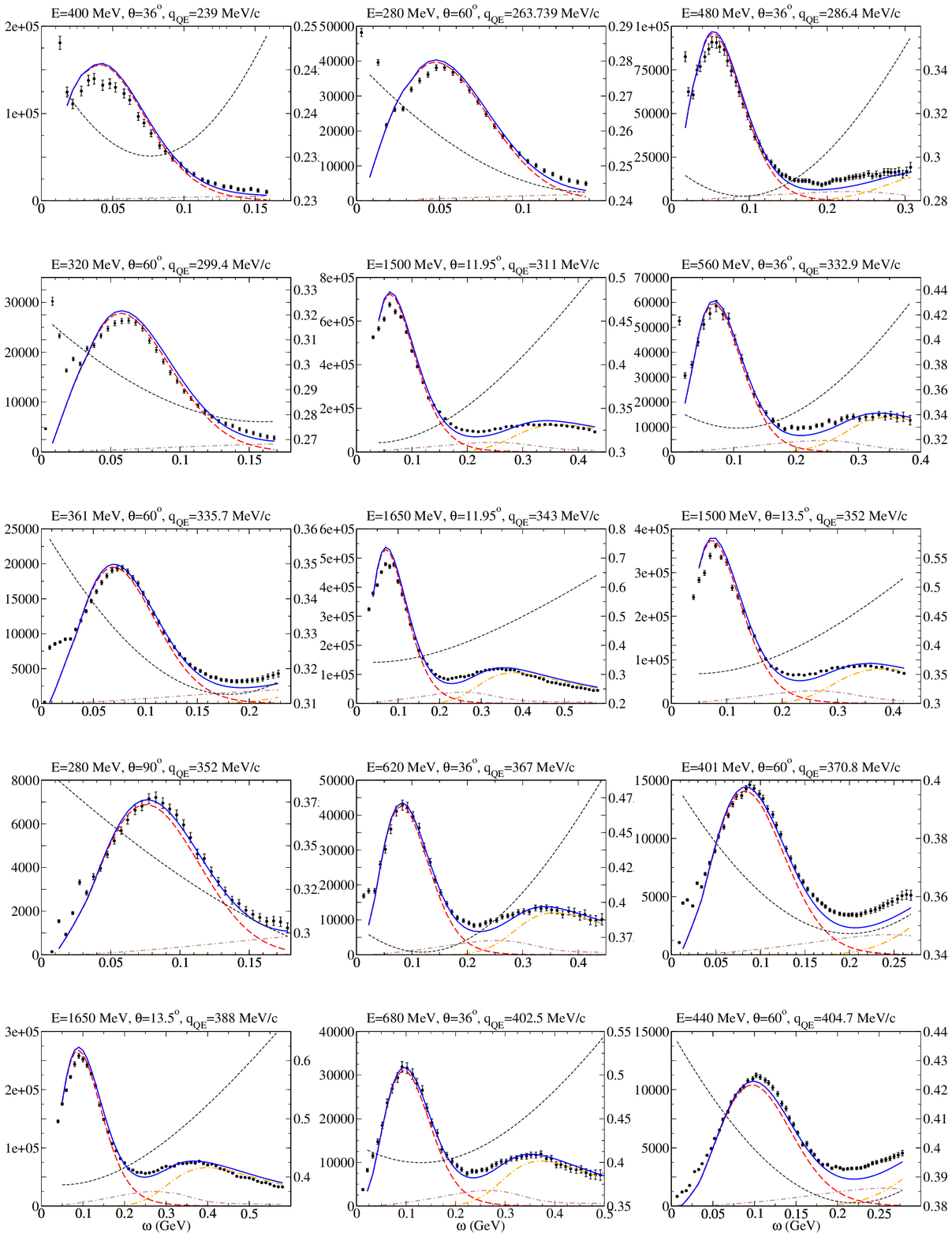}
\begin{center}
\vspace{-1cm}
\end{center}
\end{center}
\caption{(Color online) Comparison of inclusive $^{12}$C(e,e') cross sections and predictions of the QE-SuSAv2 model (long-dashed red line), 2p-2h MEC model (dot-dashed brown line) and
inelastic-SuSAv2 model (long dot-dashed orange line). The sum of the three contributions is represented with a solid blue line. The $q$-dependence with $\omega$ is also 
shown (short-dashed black line). The y-axis on the left represents $d^2\sigma/d\Omega/d\omega$ in nb/GeV/sr, whereas the one on the right represents the $q$ value 
in GeV/c.}\label{ee1}
\end{figure}

\begin{figure}[H]
\begin{center}\vspace{-1.9cm}
\includegraphics[scale=0.99, bb=50 109 594 800, clip]{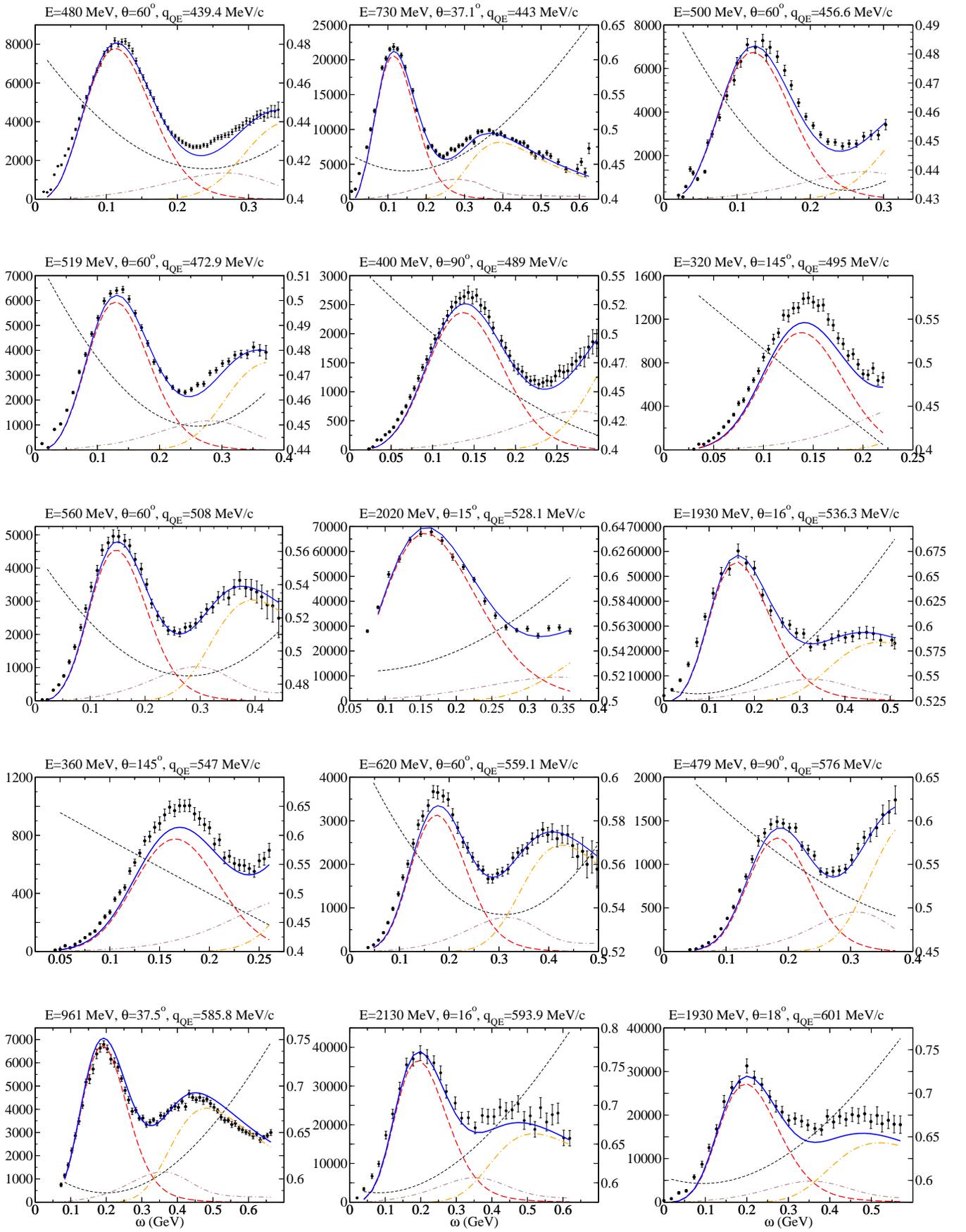}
\begin{center}
\vspace{-1cm}
\end{center}
\end{center}
\caption{(Color online) As for Fig.~\ref{ee1}, but now for kinematics corresponding to higher $q_{QE}$-values.
}\label{ee2}
\end{figure}

\begin{figure}[H]
\begin{center}\vspace{-1.9cm}
\includegraphics[scale=0.99, bb=50 109 594 800, clip]{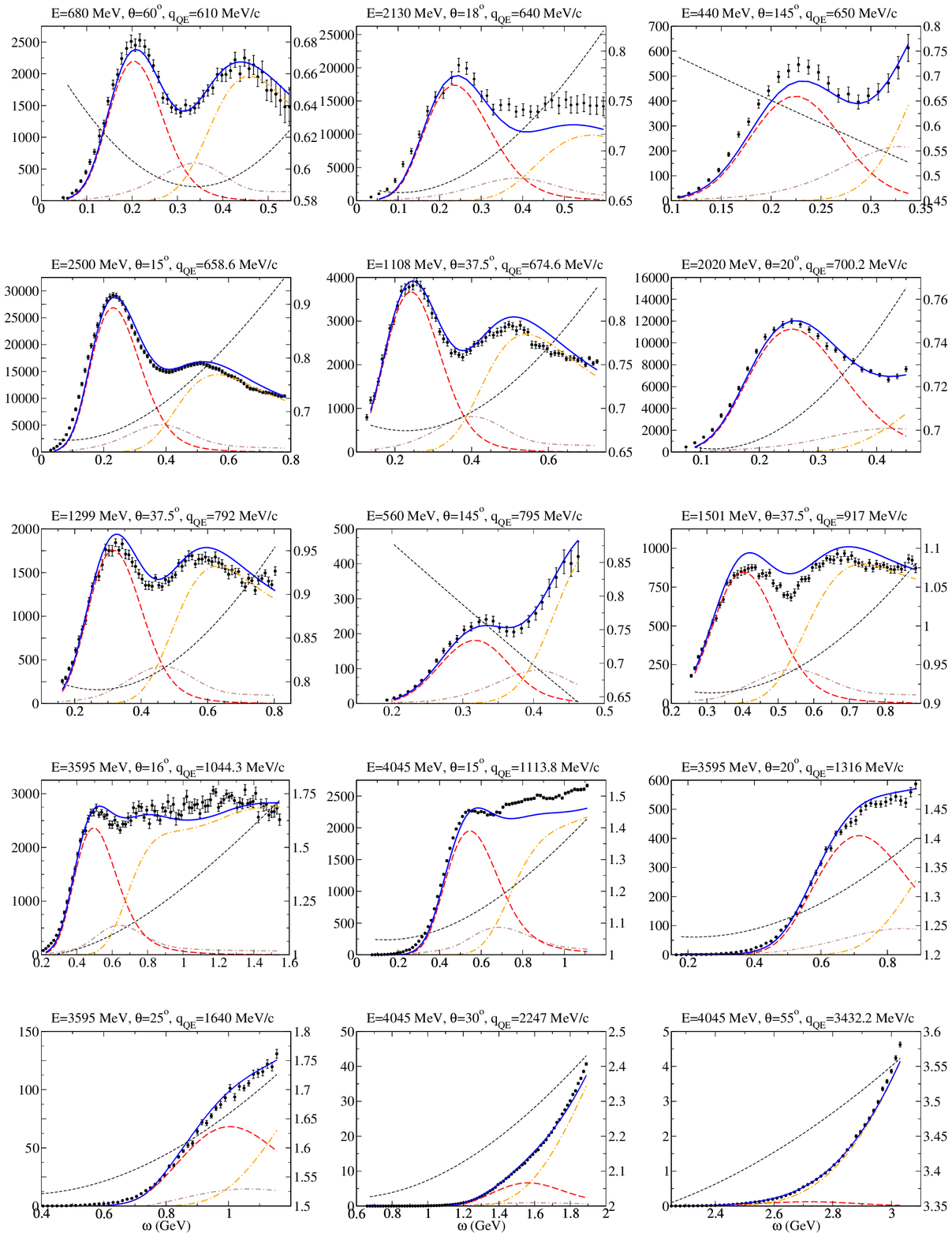}
\begin{center}
\vspace{-1cm}
\end{center}
\end{center}
\caption{(Color online) As for Fig.~\ref{ee1}, but now for kinematics corresponding to the highest $q_{QE}$-values considered.
}\label{ee3}
\end{figure}

In this section we present the double differential inclusive $^{12}$C($e,e'$) cross section versus the energy transferred to the nucleus 
($\omega$), confronting our predictions with the available experimental data~\cite{QESarchive,QESarxiv}. Results are shown in 
Figs.~\ref{ee1},~\ref{ee2} and~\ref{ee3}: in each panel we show the three separate  
contributions to the inclusive cross section, namely, QE, 2p-2h MEC and inelastic. The comparisons are carried out for a very wide range of kinematics 
from low-intermediate energies to the highly-inelastic regime. Each panel corresponds to fixed values of the 
incident electron energy ($E_i$) and the scattering angle ($\theta_e$): $E_i:280-4045$ MeV and $\theta_e:12^o-145^o$. To make it easier to discuss the results to follow, the ordering of the panels has been done according to the corresponding value for the momentum transfer at the quasielastic peak, denoted 
as $q_{QE}$. This gives us the value of $q$ where the maximum in the QE peak appears. However, it is important to point out that as $\omega$ varies, $q$ also varies. 
This is important in order to estimate the value of the RMF/RPWIA transition parameter $q_0$ in both
regimes, QE and inelastic. Hence we also include in each panel a curve that shows  how the momentum transfer changes with $\omega$.
Results illustrate that at very forward angles the value of $q$ increases with the energy transfer, whereas this trend tends to reverse at backward 
angles. Thus for electrons scattered backwards, the $q$-values corresponding to the inelastic process are smaller than those ascribed to the
QE regime. However, notice that in this situation the cross section is clearly dominated by the QE peak. On the contrary, at very forward kinematics the
inelastic process takes place at larger values of $q$. Thus, the two regimes, QE and inelastic, overlap strongly, the inelastic processes being the main ones 
responsible for the large cross sections observed at increasing values of $\omega$. Finally, for intermediate scattering angles the behavior of $q$ exhibits
a region where it decreases (QE-dominated process), whereas for higher $\omega$ (inelastic regime) the behavior of $q$ reverses and starts to go up. In these
situations the QE peak, although significantly overlapped with the inelastic contributions, is clealy visible even for very high electron energies.

The systematic analysis presented in Figs.~\ref{ee1},~\ref{ee2} and~\ref{ee3} demonstrates that the present SuSAv2-MEC model provides a very successful description
of the whole set of $(e,e')$ data, validating the reliability of our predictions. 
The positions, widths and heights of the QE peak are nicely reproduced by the model taking into account not only the QE domain but also the contributions given by
the 2p-2h MEC terms (around $\sim10-15\%$). Only at very particular kinematics, {\it i.e.,} $\theta_e=145^o$ and $E=320$ ($360$) MeV (Fig.~\ref{ee2}) and 
$440$ MeV (Fig.~\ref{ee3}), does the model clearly underpredict data at the QE peak as also observed in \cite{Martini:ee}. However, notice that the dip region is successfully
reproduced by the theory. Moreover, the remaining kinematics corresponding to very backward angles, $E=560$ MeV, $\theta_e=145^o$
(Fig.~\ref{ee3}), is well described by the model with a very high tail ascribed to the inelastic processes. Another kinematical situation whose
discussion can be of interest concerns the scattering angle $\theta_e=37.5^o$. Four cases are shown, one in Fig.~\ref{ee2} and three in Fig.~\ref{ee3}.
As noted, the model does very well for the lower values of $q_{QE}$ starting to depart from data as $q_{QE}$ goes up. Note that this is the case at
$q_{QE}=792$ MeV/c and, particularly, at $q_{QE}=917$ MeV/c where the theoretical predictions overestimate data by $5\%$ and $10\%$, respectively, at the QE peak
as well as in the dip region where the QE and inelastic contributions overlap and 2p-2h MEC are sizeable.

This overestimation of cross
  section occurs only for the set of data of ~\cite{Sealock:1989nx}, while a good agreement is observed at
  similar scattering angles, but for lower momentum transfers, namely,
  $q_{QE}=402.5$ MeV/c (Fig.~\ref{ee1}) and $q_{QE}=443$ MeV/c
  (Fig.~\ref{ee2}), which correspond to different experimental setups.

Some comments concerning the ``dip'' region between the QE and the $\Delta$ peaks are also in order. This is the region where the QE and the inelastic
contributions overlap the most and where FSI effects that modify in a significant way the tail of the QE curve at large $\omega$-values can introduce
an important impact. Moreover, the role of the 2p-2h MEC effects is essential because its maximum contribution occurs in this region. Thus,
only a realistic calculation of these ingredients beyond the IA can describe successfully the behavior of the cross section. 

To conclude, the
accordance between theory and data in the inelastic regime, where a wide variety of effects are taken into account, also gives us a great confidence in
the reliability of our calculations. Note the excellent agreement in some situations even being aware of the limitations and particular difficulties 
in order to obtain phenomenological fits of the inelastic structure functions, and the poorer quality of some experimental data sets at these kinematics.
%



\subsection{Sensitivity of the model}

\begin{figure}[H]
\begin{center}\vspace{-1.9cm}
\includegraphics[scale=0.22, angle=270]{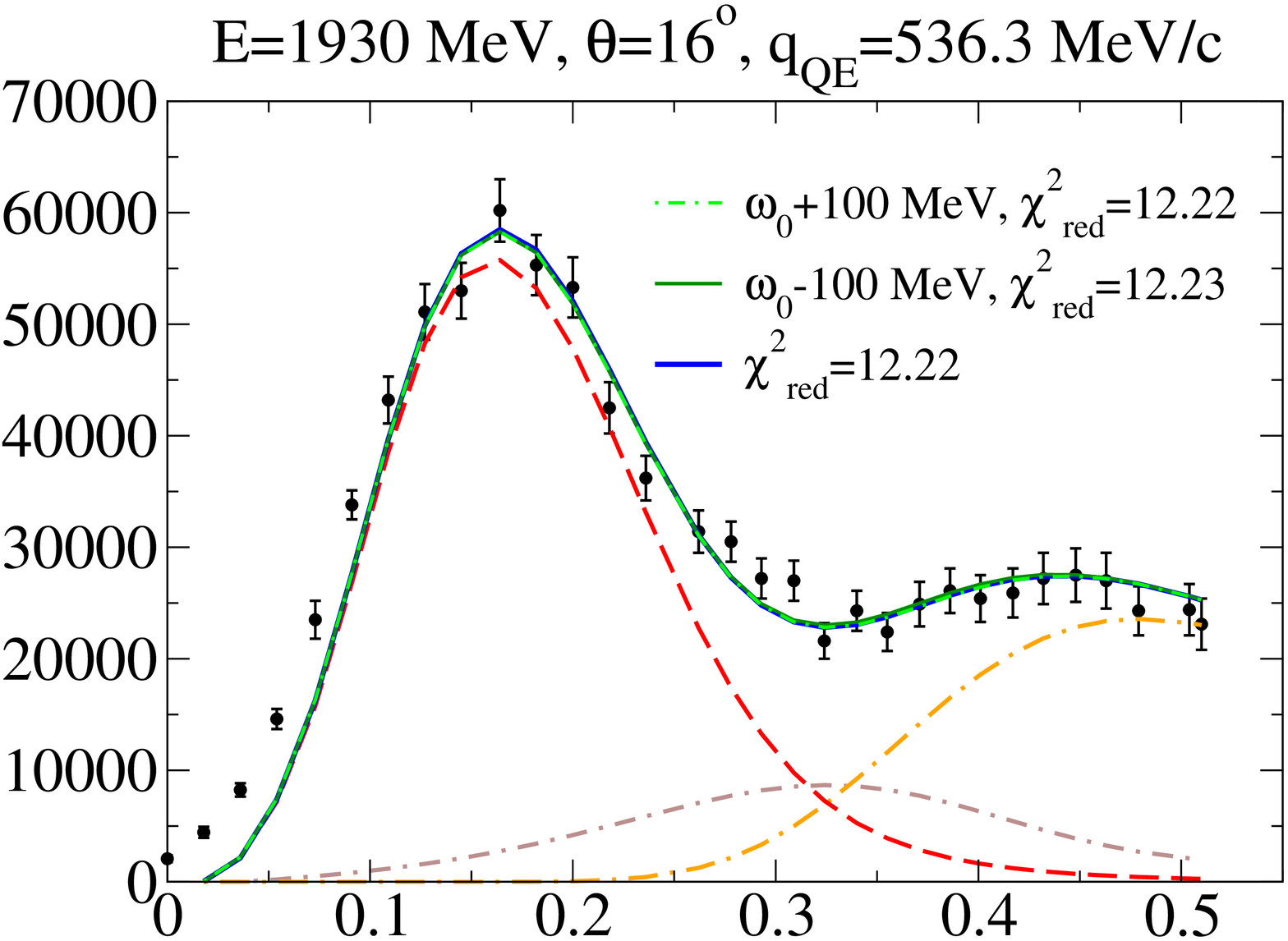}\hspace*{-0.15cm}\includegraphics[scale=0.22, angle=270]{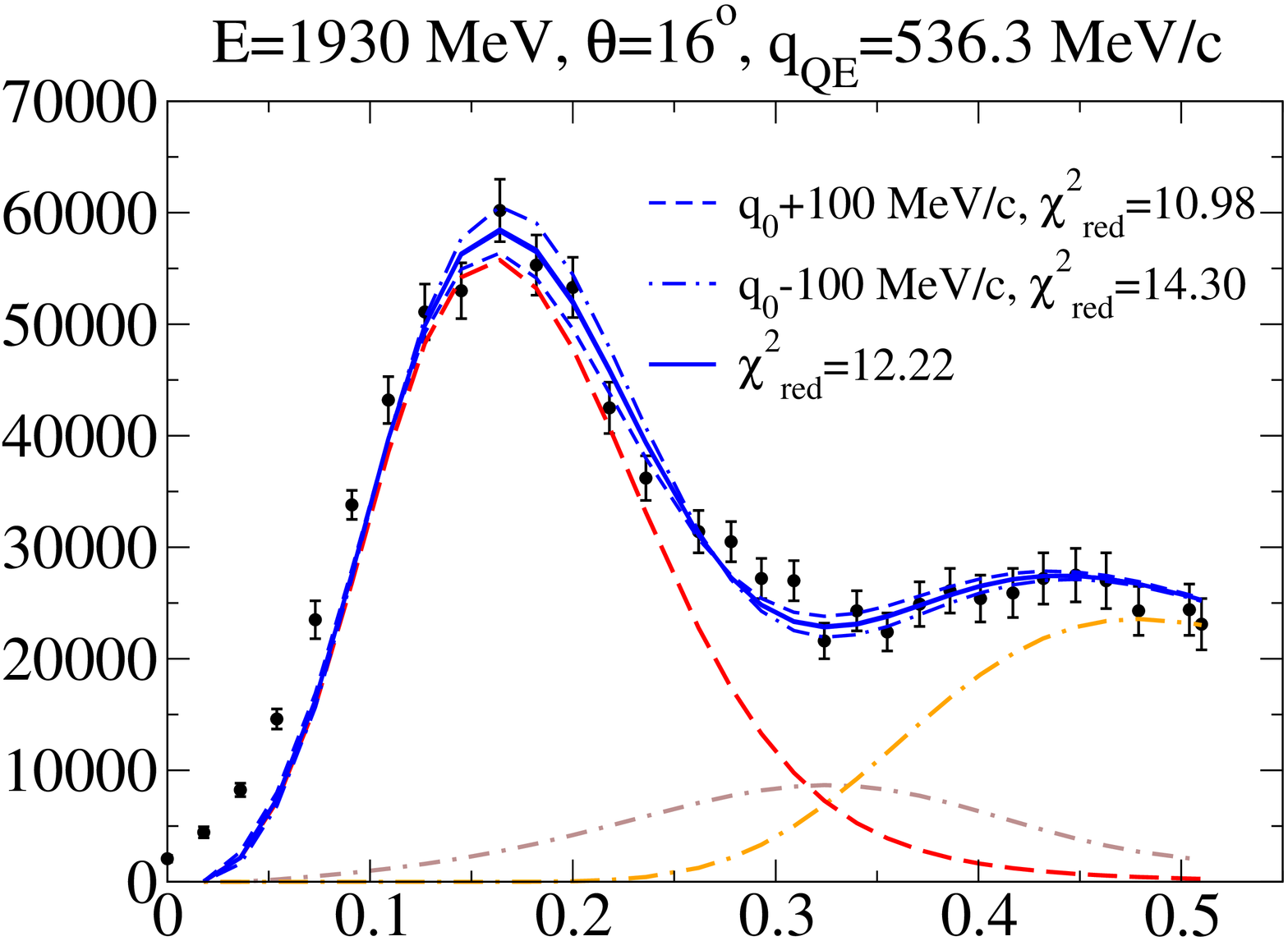}\hspace*{-0.15cm}\includegraphics[scale=0.22, angle=270]{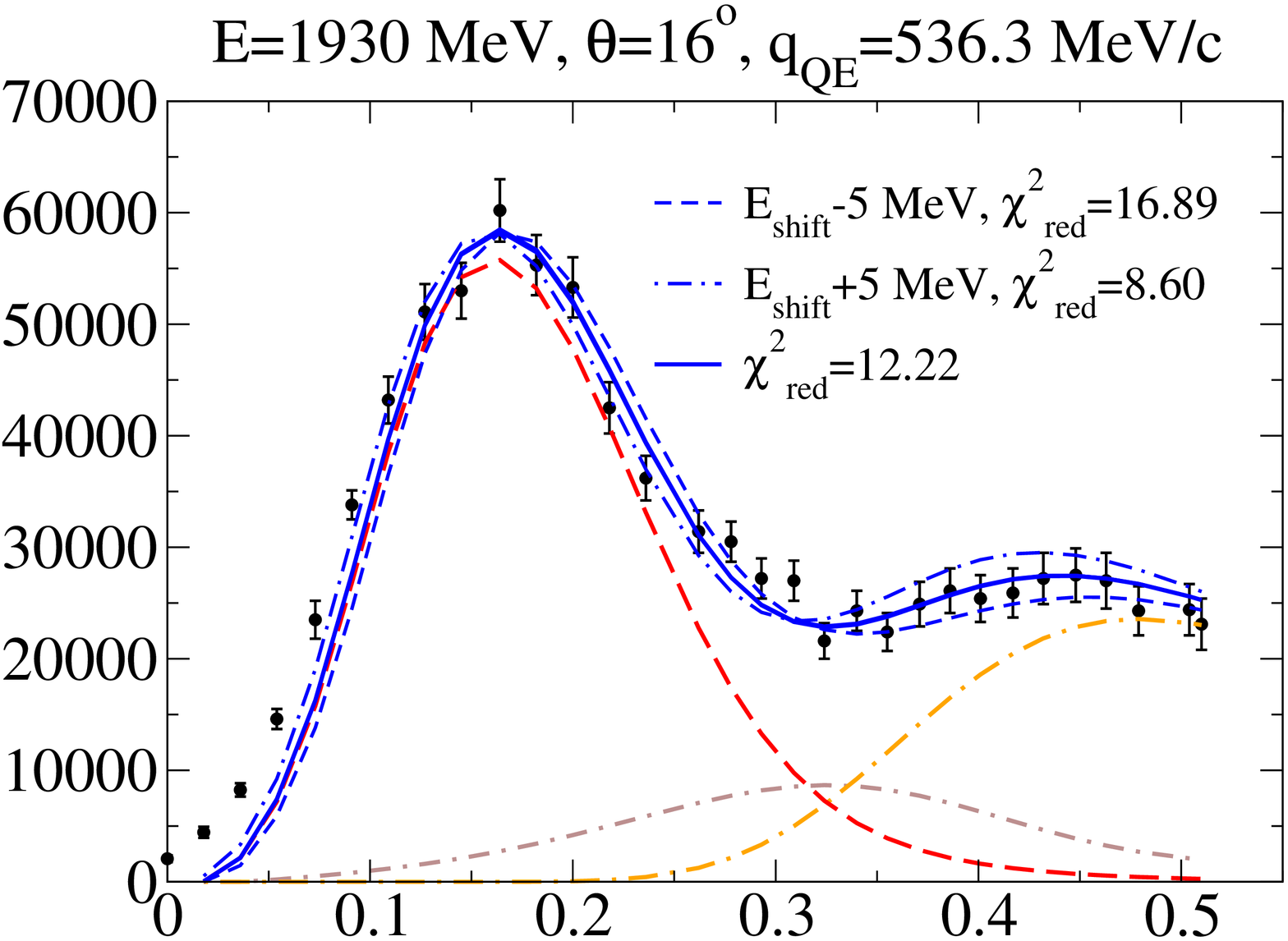}\\
\includegraphics[scale=0.22, angle=270]{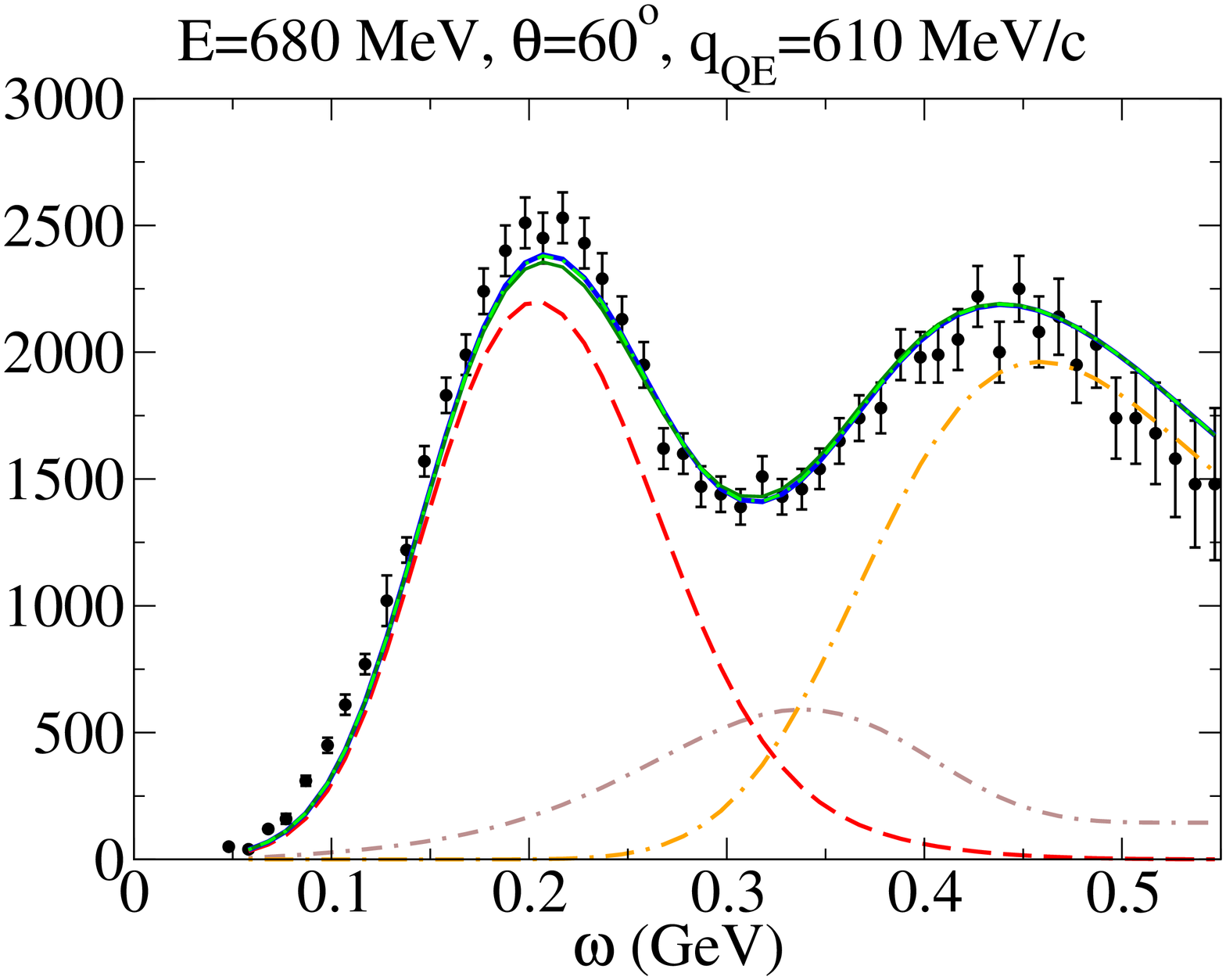}\hspace*{-0.15cm}\includegraphics[scale=0.22, angle=270]{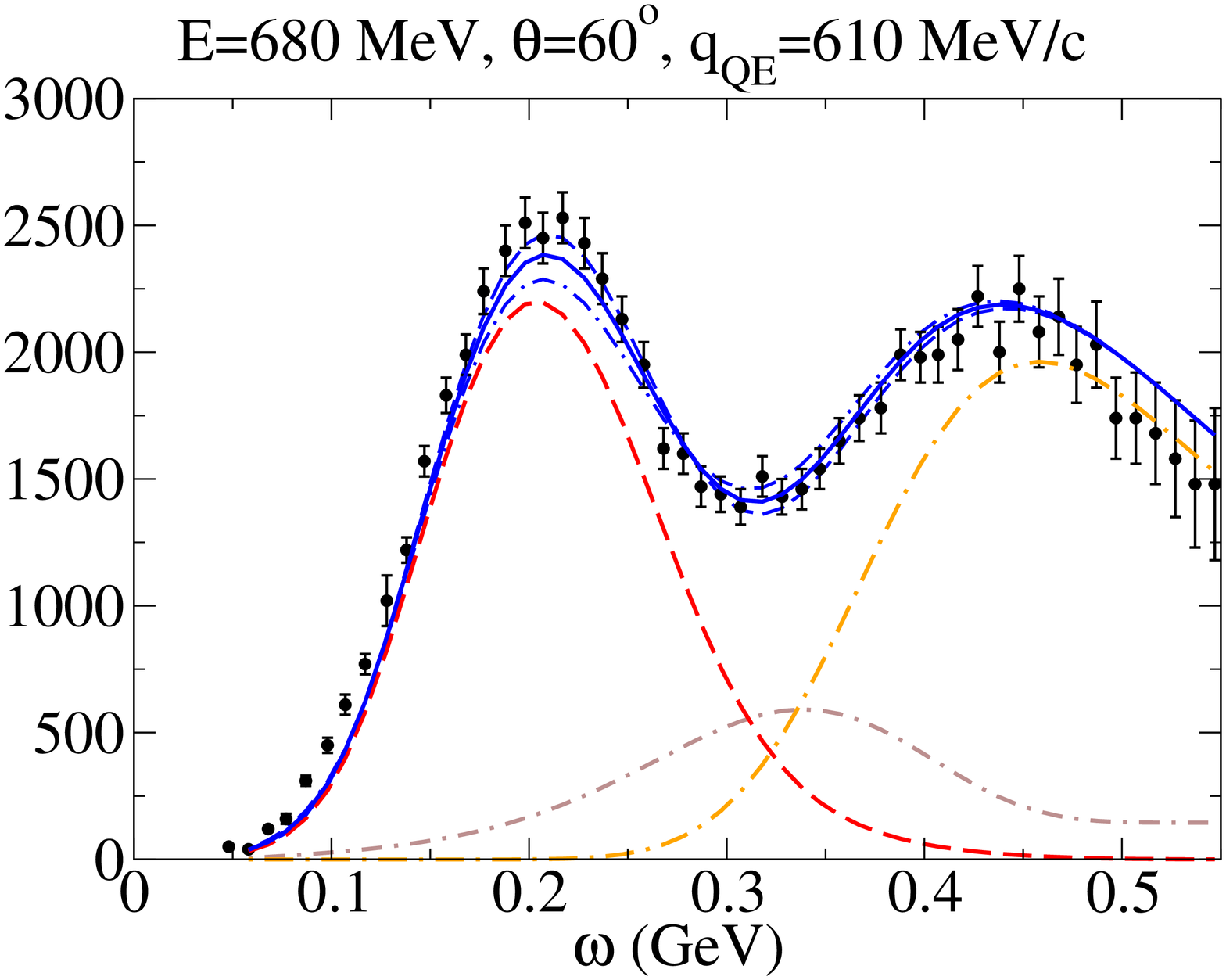}\hspace*{-0.15cm}\includegraphics[scale=0.22, angle=270]{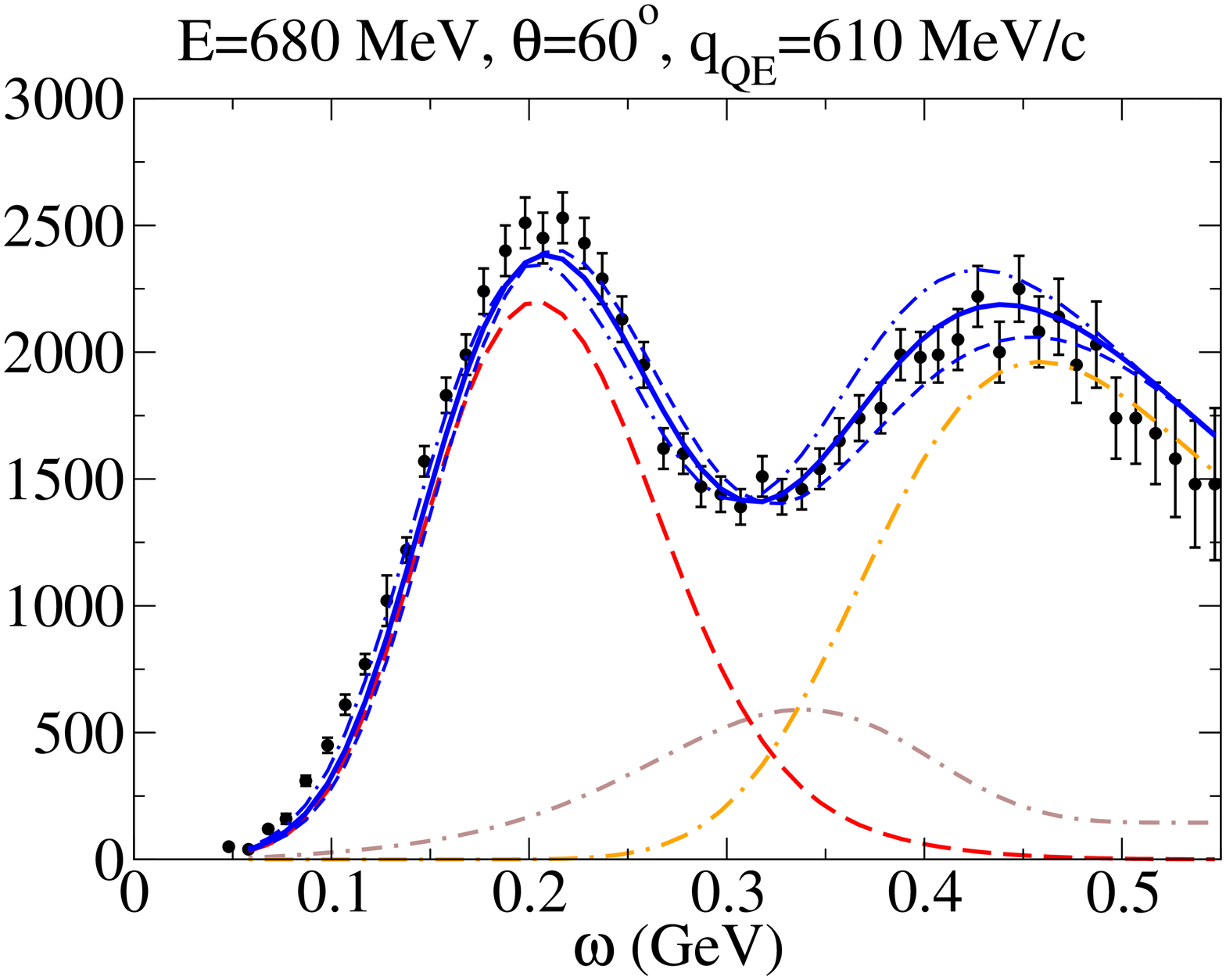}\\
\begin{center}
\vspace{-1cm}
\end{center}
\end{center}
\caption{(Color online) Comparison of inclusive $^{12}$C(e,e') cross sections and predictions of the QE-SuSAv2 model (long-dashed red line), 2p-2h MEC model (dot-dashed brown line) and
inelastic-SuSAv2 model (long dot-dashed orange line). The sum of the three contributions is represented with a solid blue line. It is also shown the total contribution by shifting $\omega_0$ (left panels), $q_0$ (middle panels) and $E_{shift}$ (right panels). The y-axis represents $d^2\sigma/d\Omega/d\omega$ in nb/GeV/sr, whereas the x-axis represents $\omega$ in GeV.}\label{ee_p}
\end{figure}

It is important to point out the novelties introduced in this work compared with some previous (preliminary) studies. With
regards to the results shown in~\cite{Maieron:2009an}, that were based only on the superscaling function extracted from the analysis
of the longitudinal $(e,e')$ data and assuming the transverse function to be equal (scaling of zeroth kind), in the present paper
the enhancement in the transverse channel introduced by the RMF model is incorporated. Moreover, the role of FSI is carefully examined
by making use of the evolution of the scaling funtion from the RMF responses to the RPWIA ones as the momentum transfer goes up. 
This explains why the present analysis provides a much more accurate description of the data. Notice that the new SuSAv2 makes both the
QE and the inelastic results higher. This outcome can be also observed in \cite{Gonzalez-Jimenez:2014eqa} where the study 
was restricted to the QE region and a fixed value of $q_0$ that can be appropriate for the specific kinematics considered was used.
On the contrary, here the aim is to provide a model capable of reproducing $(e,e')$ cross sections for a very wide selection of 
kinematics and including in each case the whole energy spectrum. This is consistent with the $q$-dependence shown by $q_0$ in both
regimes, QE and inelastic. We have also tested the sensitivity of our results to different
choices in the values of $\omega_0$, $q_0$ and $E_{shift}$ for two representative kinematical situations (see Fig.~\ref{ee_p}). With regards to $\omega_0$ a variation of $\pm 100$ MeV leads to negligible
effects, hence the value of $\chi^2$ is basically the same (left panels in Fig.~\ref{ee_p}). 
In the case of $q_0$ and $E_{shift}$, variations of the order of $\pm 100$ MeV/c (in $q_0$)
and $\pm 5$ MeV ($E_{shift}$) lead to differences within $\sim20\%$ on $\chi^2$, but still providing a very good representation of the data (see results presented in the middle and right panels of Fig.~\ref{ee_p}). Note however that
$q_0$ is a dynamical parameter running with $q$, whereas the value of $E_{shift}$ is determined by the right location of the maxima in the scaling
functions. Hence a significant variation of these three values does not imply a worsening in the agreement with data.

\subsection{Relevance of the RMF/RPWIA effects}

\begin{figure}[H]
\begin{center}\vspace{-1.9cm}
\includegraphics[scale=0.99, bb=50 325 594 800, clip]{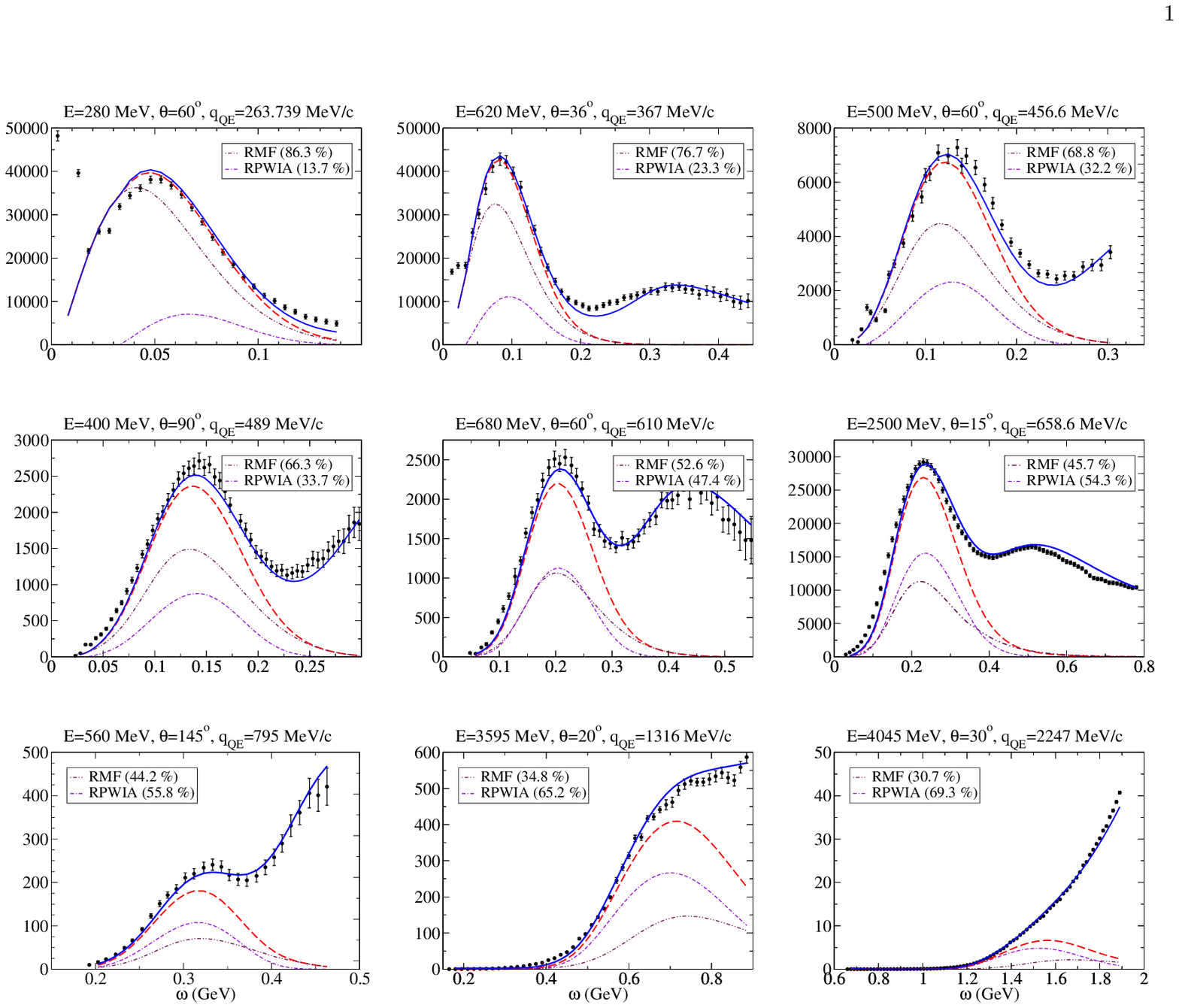}
\begin{center}
\vspace{-1cm}
\end{center}
\end{center}
\caption{(Color online) Comparison of RMF and RPWIA contributions in the QE regime. Also shown for reference the predictions of the total QE-SuSAv2 model (long-dashed red line) and the total inclusive contribution (solid blue line). The y-axis represents 
$d^2\sigma/d\Omega/d\omega$ in nb/GeV/sr.}\label{eermfrpwia}
\end{figure}

The SuSAv2 model discussed in this work incorporates ingredients coming from the RMF and RPWIA approaches. Whereas the RMF
provides an excellent description of the experimental longitudinal scaling function extracted from data taken at intermediate $q$-values, producing
the required asymmetry and the enhancement of the transverse response, the RPWIA approach yields much more suitable results at higher values of the
momentum transfer where FSI effects are significantly reduced. In Fig.~\ref{eermfrpwia} we present the cross sections for a set of kinematical
situations showing the isolated contributions emerging from the two models in the case of the QE regime. The
percentage of the two contributions is given in each panel. As shown, for those kinematics that correspond to the lower values of $q_{QE}$ (top panels) the
RMF response contributes the most. As $q_{QE}$ increases, the RPWIA contribution becomes relatively more important, approaching the RMF one (see panels in the
middle). Finally, for the higher $q_{QE}$-values (bottom panels) the behavior reverses with the RPWIA result being the main one responsible for the QE response.

\begin{figure}[H]
\begin{center}
\includegraphics[scale=0.475, angle=270]{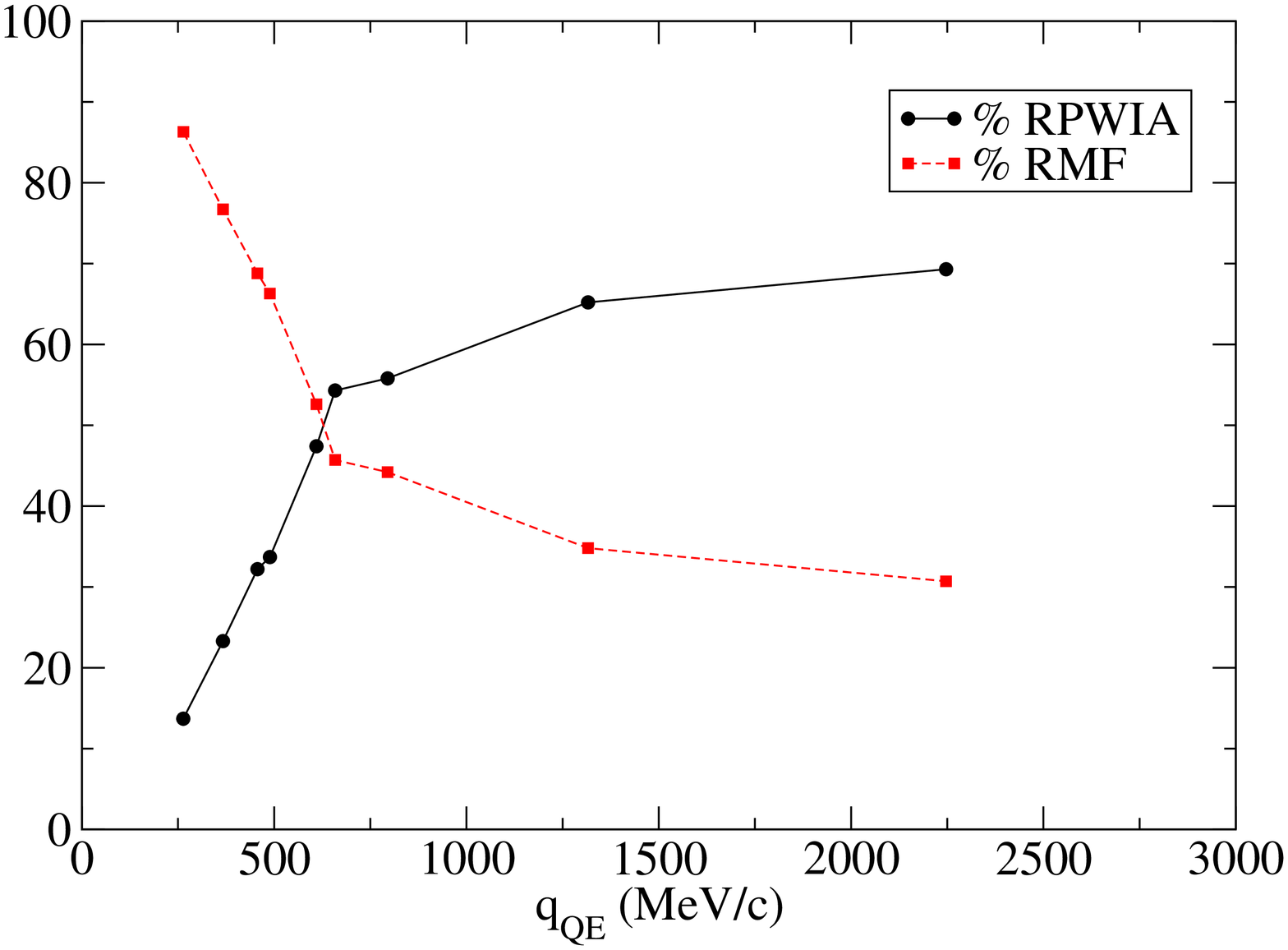}
\begin{center}
\vspace{-1cm}
\end{center}
\end{center}
\caption{(Color online) Comparison of percentages corresponding to the RMF and RPWIA contributions in the QE regime as a function of $q_{QE}$.}\label{eeperc}
\end{figure}

To make clearer how both RMF and RPWIA approaches contribute within the SuSAv2 model, in Fig.~\ref{eeperc} we present the specific percentages
ascribed to the two contributions and how they vary with the value of $q_{QE}$. The main variation in the two cases is produced in the
region of intermediate $q_{QE}$-values, namely, $250\lesssim q_{QE} \lesssim 700$ MeV/c. Here, the relative RMF contribution quickly diminishes as
$q_{QE}$ increases whereas the opposite occurs for the RPWIA. Note that at $q_{QE}\sim 700$ MeV/c both models produce basically the same answer
($\sim 50\%$) crossing each other, whereas for $q_{QE}\lesssim 500$ MeV/c RPWIA gives a very minor contribution, that is, FSI are essential to describe
data at these kinematics. Finally, at higher $q_{QE}$ the RPWIA increases slowly, whereas the RMF decreases, although in both cases some kind of saturation
seems to emerge approaching the RPWIA percentage to $\sim 60-70\%$ ($\sim 30-40\%$ for the RMF). Although not presented here for simplicity,
similar comments can be also drawn for the RMF and RPWIA contributions in the inelastic regime.

\section{Conclusions}

The SuSAv2 model was originally introduced in~\cite{Gonzalez-Jimenez:2014eqa} and applied to the analysis of electron and charged-current (CC) neutrino scattering
reactions within the QE domain, that is, the model was based exclusively on the IA. Contrary to the original SuSA model, based on
the existence of a universal scaling function extracted from the longitudinal $(e,e')$ data, the SuSAv2 model incorporates several ``reference'' scaling
functions related to the predictions given by the RMF approach. This leads to zeroth-kind scaling violations, namely, the transverse scaling function
is higher by $\sim 20\%$ than the longitudinal one. Furthermore, the difference between isoscalar and isovector contributions in electron
and neutrino reactions as well as the axial-axial and the interference axial-vector terms in the latter, introduce basic differences that
are incorporated in the new SuSAv2. All these ingredients have been taken into account in addition to the particular behavior shown by the scaling
functions versus the momentum/energy transfer in the process. Whereas the RMF approach does well at low to intermediate values of $q$, results in the
high-$q$ regime revert to those of the RPWIA. Hence SuSAv2 is constructed as a ``blend'' between the properties of the RMF and RPWIA approaches.

In this work the SuSAv2 model is extended for the first time to the whole energy spectrum, incorporating the contributions coming from the QE,
inelastic and two-body meson exchange currents. Within this framework a general ``blending'' function is introduced to make the transition
between the RMF and RPWIA responses. This function
is constructed in terms of a parametrization of the optimized blending region given by $q_0$, and it has been applied to the QE as well as to the inelastic regimes.  Although the use of more free parameters, as $\omega_0$ and/or 
the shift energy, leads to an even better agreement with data in some particular cases, the specific parametrization assumed is not critical,
and indeed, the present model is capable of reproducing very successfully the whole energy spectrum of $(e,e')$ data at very different kinematics.
This gives us a great confidence in the reliability of the model when extended to the description of neutrino-nucleus scattering. In this case,
not only new responses contribute, but also the wide neutrino energy band implied by the typical accelerator-based neutrino fluxes makes it difficult to reconstruct the
neutrino energy. Thus, ingredients beyond the ones usually assumed within the IA can have a significant impact on the analysis of data. Work along
these lines is presently in progress.

A basic feature of our present study, apart from the SuSAv2 model
applied to the QE and inelastic regions, concerns the evaluation of
the two-body meson exchange currents. This is based on a fully
relativistic model that can be thus applied to very high
energies/momenta. This is crucial in order to analyze neutrino
oscillation experiments.  
In the present study we have used a fixed parameterization of the 2p-2h MEC response functions that allows us to avoid the computationally demanding microscopic calculation for the entire set of kinematics for the experimental data presented here, including for the first time both the transverse and longitudinal two-body currents.

In future work we will present a similar parametrization 
of the 2p-2h MEC responses for use in CC neutrino scattering (see~\cite{Simo2016ikv}).
This can be easily incorporated into the Monte Carlo
  neutrino event generators, and this should be of great
  interest for analyses of neutrino oscillation experiments.  

To conclude, we emphasize the importance of scaling arguments and the need to describe properly electron scattering
data before the model can be extended to neutrino reactions. The analysis presented in this work, restricted to electrons, includes
a complete relativistic calculation of the MEC-2p2h contributions in addition to the global scaling analysis applied not only to the
QE regime but also to the inelastic one. These ingredients are of crucial importance for the analysis of neutrino reactions.

\begin{acknowledgments}

 This work was partially supported by INFN under
 project MANYBODY, by Spanish Direccion General de Investigacion
 Cientifica y Tecnica and FEDER funds (grant No. FIS2014-53448-C2-1
 and grant No. FIS2014-59386-P), by the Junta de Andalucia (grant
 No. FQM225), by the Spanish Consolider-Ingenio 2000 program CPAN
 (CSD2007-00042), by Andalucia Tech and by the Office of Nuclear Physics of the US Department of Energy under Grant Contract Number DE-FG02-94ER40818 (TWD). GDM acknowledges support from a
 fellowship from the Junta de Andalucia (FQM-7632, Proyectos de
 Excelencia 2011). The authors would like to thank Arturo De Pace, Raul Gonz\'alez-Jim\'enez and Ignacio Ruiz-Sim\'o for interesting discussions.
\end{acknowledgments}


\bibliography{iopart-num}

\end{document}